# On the Behaviour of Non–Singlet Structure Functions at Small $x$

J. Blümlein$^a$ and A. Vogt$^{b,1}$

$^a$*DESY–Zeuthen*
*Platanenallee 6, D–15735 Zeuthen, Germany*

$^b$*Deutsches Elektronen-Synchrotron DESY*
*Notkestraße 85, D–22603 Hamburg, Germany*

**Abstract**

The resummation of $O(\alpha_s^{l+1} \ln^{2l} x)$ terms in the evolution kernels of non–singlet combinations of unpolarized and polarized structure functions is investigated. The agreement with complete calculations up to order $\alpha_s^2$ is demonstrated, and the leading small-$x$ contributions to the three–loop non–singlet splitting functions $P^{\pm}$ are derived. The additional contributions due to the resummed terms are studied numerically for the most important non–singlet structure functions. They are found to be about 1 % or smaller in the kinematical regions accessible at present and in the foreseeable future.

---

$^1$On leave of absence from Sektion Physik, Universität München, D-80333 Munich, Germany

# 1 Introduction

The resummation of leading contributions in the evolution kernels of singlet structure function combinations at small $x$ [1] may lead to large effects [2]. In this case, the small-$x$ evolution is dominated by the rightmost singularity in the $N$-moment plane $\sim (\alpha_s/[N-1])^l$ and higher order $\alpha_s$ corrections to it. Such terms are absent in the non-singlet kernels both for the unpolarized and polarized structure functions [3, 4], as well as in the singlet kernels in the polarized case [5, 6]. Here the most singular contributions behave like $N(\alpha_s/N^2)^l$. An all-order resummation of these terms for non-singlet structure functions has been worked out in ref. [7]. Very sizeable corrections due to this resummation have been claimed for both unpolarized and polarized structure functions recently [8]. In this way the small-$x$ behaviour of the structure function evolution, e.g. of $xF_3^{\nu d}(x,Q^2)$, $F_2^p(x,Q^2) - F_2^n(x,Q^2)$, and $g_1^p(x,Q^2) - g_1^n(x,Q^2)$, may be considerably affected.

So far the resummation [7] was compared with the results of complete calculations only in the universal term of order $\alpha_s/N$ [7, 8]. After setting up our notation and recalling the standard NLO formulation in section 2, we will show in section 3 that the resummation [7] agrees with the known evolution kernels $P_{\rm NS}^{\pm}(x, \alpha_s)$ in next to leading order (NLO) for $q^2 < 0$ as well in the small-$x$ limit. The contributions $\propto \alpha_s^3 \ln^4 x$ to the so far uncalculated 3-loop non-singlet $\overline{\rm MS}$ splitting functions are then derived from the results of ref. [7]. In section 4, we perform a numerical analysis for the most important non-singlet structure functions and compare the effect of the new terms beyond next to leading order with the NLO results.

# 2 Evolution in fixed-order perturbative QCD

The evolution equation for the non-singlet combinations $f_{\rm NS}^{\pm}(x, Q^2)$ of parton densities reads

$$\frac{\partial f_{\rm NS}^{\pm}(x, Q^2)}{\partial \ln Q^2} = P_{\rm NS}^{\pm}(x, \alpha_s) \otimes f_{\rm NS}^{\pm}(x, Q^2) \ . \tag{1}$$

Here $\otimes$ denotes the Mellin convolution, and $P_{\rm NS}^{\pm}(x, \alpha_s)$ is specified below. In the following, all our expressions refer to the $\overline{\rm MS}$ factorization and renormalization scheme, and we drop the subscript 'NS' wherever the non-singlet character of the quantity under consideration is obvious from the superscript '$\pm$'. The splitting function combinations $P^{\pm}(x, \alpha_s)$ are given by

$$P^{\pm}(x, \alpha_s) = P_{qq}(x, \alpha_s) \pm P_{q\bar{q}}(x, \alpha_s) \equiv \sum_{l=0}^{\infty} a_s^{l+1} P_l^{\pm}(x) \tag{2}$$

with $a_s \equiv \alpha_s(Q^2)/(4\pi)$. We now restrict ourselves to the spacelike case, $Q^2 = -q^2 > 0$. The expansion coefficients $P_l^-(x)$ obey the sum rule

$$\int_0^1 dx P_l^-(x) = 0 \qquad \forall \ l \ \epsilon \ \boldsymbol{N} \ , \tag{3}$$

which is due to fermion number conservation and Weierstrass' theorem, since $a_s$ acts as an independent parameter. At present the splitting functions are known up to two-loop order [3, 4]. They read

$$\begin{aligned} P_{qq}(x, a_s) &= 2a_s C_F \left[\frac{1+x^2}{1-x}\right]_+ \\ &\quad + 4a_s^2 \left[C_F^2 P_F(x) + \frac{1}{2} C_F C_G P_G(x) + C_F N_f T_R P_{N_f}(x)\right] + \mathcal{O}(a_s^3) \end{aligned} \tag{4}$$

$$P_{q\bar{q}}(x, a_s) = 4a_s^2 \left[C_F^2 - \frac{1}{2} C_F C_G\right] P_A(x) + \mathcal{O}(a_s^3) \ , \tag{5}$$

where $C_F = (N_c^2 - 1)/(2N_c), C_A = N_c$, and $T_R = 1/2$. $N_f$ denotes the number of active flavours. The functions $P_I(x)$, $I = F, G, N_f, A$ were derived in refs. [4]. For $x \to 0$ the leading contributions to $P^{\pm}(x, a_s)$ are

$$P^+_{x \to 0}(x, a_s) = 2a_s C_F + 2a_s^2 C_F^2 \ln^2 x + \mathcal{O}(a_s^3)$$
$$P^-_{x \to 0}(x, a_s) = 2a_s C_F + 2a_s^2 \left[-3C_F^2 + 2C_F C_G\right] \ln^2 x + \mathcal{O}(a_s^3) \ . \tag{6}$$

The (scheme dependent) parton densities $f^{\pm}(x, Q^2)$ are no observables themselves beyond the leading order. Instead of their evolution equation (1), one can directly consider the evolution of the structure functions $F_i^{\pm}(x, Q^2)$ obtained by the convolution

$$F_i^{\pm}(x, Q^2) = c_i^{\pm}(x, Q^2) \otimes f_i^{\pm}(x, Q^2) \ . \tag{7}$$

Here $c_i^{\pm}(x, Q^2)$ denote the coefficient functions

$$c_i^{\pm}(x, Q^2) = \delta(1-x) + \sum_{l=1}^{\infty} a_s^l c_{i,l}^{\pm}(x) \tag{8}$$

corresponding to $F_i^{\pm}(x, Q^2)$. The evolution equation for $F_i^{\pm}$ can be rewritten as an equation in $a_s(Q^2)$ rather than in $Q^2$ using

$$\frac{\partial a_s}{\partial \ln Q^2} = -\beta_0 a_s^2 - \beta_1 a_s^3 + \mathcal{O}(a_s^4) \ . \tag{9}$$

This leads to

$$\frac{\partial F_i^{\pm}(x, a_s)}{\partial a_s} = -\frac{1}{\beta_0 a_s^2} K_i^{\pm}(x, a_s) \otimes F_i^{\pm}(x, a_s) \ , \tag{10}$$

where the NLO evolution kernels $K_{i,1}^{\pm}$ are given by

$$K_{i,1}^{\pm}(x, a_s) = P_{\text{NS},0}(x) a_s + \left[P_1^{\pm}(x) - \frac{\beta_1}{\beta_0} P_{\text{NS},0}(x) - \beta_0 c_{i,1}^{\pm}(x)\right] a_s^2 \ . \tag{11}$$

Note the obvious fact that the $\ln^2 x$ terms of eqs. (6) enter the evolution equation (10) only in combination with the coefficient $\beta_0$ in eq. (9).

## 3 Resummation of leading small-$x$ terms

The transformation to Mellin-$N$ space

$$\mathcal{M}\left[K^{\pm}_{x \to 0}(a_s)\right](N) = \int_0^1 dx \, x^{N-1} K^{\pm}_{x \to 0}(x, a_s) \equiv -\frac{1}{2} \Gamma^{\pm}_{x \to 0}(N, a_s) \tag{12}$$

of the most singular part of the evolution kernels $K^{\pm}$ in all orders in $a_s$ was given in ref. [7][2]:

$$\Gamma^+_{x \to 0}(N, a_s) = -N \left\{ 1 - \sqrt{1 - \frac{8a_s C_F}{N^2}} \right\}$$

$$\Gamma^-_{x \to 0}(N, a_s) = -N \left\{ 1 - \sqrt{1 - \frac{8a_s C_F}{N^2} \left[1 - \frac{8N_c a_s}{N} \frac{d}{dN} \ln\left(e^{z^2/4} D_{-1/[2N_c^2]}(z)\right)\right]} \right\} \ . \tag{13}$$

---
[2]Note that there are a few misprints in eq. (4.7) of ref. [7].

Here $z = N/\sqrt{2N_c a_s}$, and $D_p(z)$ denotes the function of the parabolic cylinder [9]. We expand eqs. (13) into a power series in $a_s^k/N^{2k-1}$, and transform the result to $x$-space using

$$\mathcal{M}\left[\ln^k\left(\frac{1}{x}\right)\right](N) = \frac{k!}{N^{k+1}} . \tag{14}$$

One obtains

$$K_{x\to 0}^+(x, a_s) = 2a_s C_F + 2a_s^2 C_F^2 \ln^2 x + \frac{2}{3} a_s^3 C_F^3 \ln^4 x + \mathcal{O}(a_s^4 \ln^6 x) \tag{15}$$

$$K_{x\to 0}^-(x, a_s) = 2a_s C_F + 2a_s^2 C_F \left[C_F + \frac{2}{N_c}\right] \ln^2 x + \frac{2}{3} a_s^3 C_F \left[C_F^2 - \frac{3}{2N_c^2}\right] + \mathcal{O}(\alpha_s^4 \ln^6 x) .$$

The expressions (15) agree with the corresponding result found for $P_{x\to 0}^\pm(x, a_s)$, eqs. (6), in the complete NLO calculations of the non–singlet anomalous dimensions [4] in the most singular terms since

$$C_G - \frac{3}{2} C_F = \frac{1}{N_c} + \frac{1}{2} C_F \tag{16}$$

holds in $SU(N_c)$.

Besides the terms due to the anomalous dimensions $P_l^\pm(x)$, also the coefficient functions $c_{i,l}^\pm(x)$ contribute in the evolution equation (10). The latter quantities have been calculated to $\mathcal{O}(a_s^2)$ for the structure functions $F_2(x, Q^2)$, $xF_3(x, Q^2)$ and $g_1(x, Q^2)$ [10, 11]. Expanding the coefficient functions for $x \to 0$, one finds that[3]

$$c_{i,1}(x) \propto \ln\left(\frac{1}{x}\right) \tag{17}$$

$$c_{i,2}(x) \propto \ln^3\left(\frac{1}{x}\right) . \tag{18}$$

Therefore the terms of $\mathcal{O}(a_s^2)$ and $\mathcal{O}(a_s^3)$ in eqs. (15) can be *identified* with the parts of the non–singlet anomalous dimensions proportional to $a_s(a_s \ln^2 x)^l$, assuming the validity of the resummation performed in ref. [7]. These contributions to $P_2^\pm(x)$ read

$$P_{2, x\to 0}^+(x) = \frac{2}{3} C_F^3 \ln^4 x$$

$$P_{2, x\to 0}^-(x) = \left(-\frac{10}{3} C_F^3 + 4 C_F^2 C_G - C_F C_G^2\right) \ln^4 x . \tag{19}$$

The calculation of the complete kernels $K^\pm(x, a_s)$ in eq. (10) in higher orders in $a_s$ requires to take into account also higher orders in the $\beta$–function. However, like in the NLO evoluOCon equation of section 2, the leading small-$x$ terms in $K^\pm \propto a_s(a_s \ln^2 x)^l$ do not occur together with factors containing the coefficients $\beta_i|_{i \geq 1}$.

It should be stressed that the agreement of the NLO terms between eqs. (15) obtained from the resummation [7] and eqs. (6) holds for $q^2 < 0$ only. For the time–like case $q^2 > 0$ one has [4][4]

$$P_{x\to 0}^+(x, a_s) = 2a_s C_F - 2a_s^2 C_F^2 \ln^2 x + \mathcal{O}(a_s^3)$$

$$P_{x\to 0}^-(x, a_s) = 2a_s C_F + 2a_s^2 \left[-5 C_F^2 + 2 C_F C_G\right] \ln^2 x + \mathcal{O}(a_s^3) . \tag{20}$$

---

[3]Note that apparent terms $\propto 1/x^m, m = 1, 2$ cancel in the corresponding expressions of ref. [11].

[4]The $\mathcal{O}(\alpha_s)$ coefficient functions for $e^+ e^-$ annihilation or the Drell–Yan process behave also at most $\propto \ln(1/x)$, see ref. [10].

The difference between eqs. (6) and (20) is due to the violation of the Gribov–Lipatov relation in the $\ln^2 x$ term of the NLO splitting functions.

Considerations similar to those of ref. [7] may be valid for the spacelike twist-2 singlet evolution equations in the polarized case, where the matrix of the splitting functions up to two loops has been obtained in refs. [5, 6].

## 4 Numerical results

In moment space the evolution equation (10) for the non–singlet structure functions reduces to an ordinary differential equation. Taking into account the resummed kernels (13), the solution reads

$$F^{\pm}(N, a_s) = F^{\pm}(N, a_0) \left(\frac{a_s}{a_0}\right)^{\gamma_{\text{NS},0}(N)/2\beta_0} \qquad (21)$$
$$\times \left\{\exp\left[\frac{1}{2\beta_0}\int_{a_0}^{a_s} da\, \frac{1}{a^2}\Gamma^{\pm}(N, a_s)\right] + \frac{a_s - a_0}{2\beta_0}\left[\tilde{\gamma}_1^{\pm}(N) - \frac{\beta_1}{2\beta_0}\gamma_{\text{NS},0}(N) + 2\beta_0 \hat{c}_{i,1}(N)\right]\right\}$$

with

$$\gamma_i^{\pm}(N) = -2\int_0^1 dx\, x^{N-1} P_i^{\pm}(x)\,, \quad \hat{c}_i^{\pm}(N) = \int_0^1 dx\, x^{N-1} c_i^{\pm}(x) \qquad (22)$$

and $a_0 = a_s(Q_0^2)$. Here $\tilde{\gamma}_1^{\pm}(N)$ denotes the two–loop anomalous dimension with the $1/N^3$ term subtracted, since this contribution is accounted for already in the exponential factor. Moreover,

$$\Gamma^{\pm}(N, a_s) = \mathbf{\Gamma}_{x\to 0}^{\pm}(N, a_s) + \frac{a_s}{N} C_F\,. \qquad (23)$$

The well–known solution in NLO for the evolution of $F^{\pm}(N, a_s)$ can be recovered from (21) by expanding the exponential to order $a_s$.

In the case of the non–singlet '+'-combinations the remaining integral in (21) can be done analytically, whereas it has to be performed numerically for the '−'-combinations involving the parabolic cylinder function. The transformation of the solution back into $x$–space finally affords one standard numerical integral in the complex $N$-plane [12]. We have also expanded the functions $\Gamma^{\pm}(N, a_s)$ in the coupling constant $a_s$. We find that in the practical cases considered below, one gets more than 90% of the resummation effect from the first two terms of the $\alpha_s$ expansion.

As it stands, eq. (21) violates the fermion number conservation for the '−' non–singlet combinations. Here the conjecture is that the coefficient functions $c_{i,l}^{\pm}(x)$ *do not* contain terms $\propto \ln^{2l} x$. For this no proof exists yet, however, we have verified this behaviour up to 2–loop order in section 3 for the coefficient functions of $xF_3$, $F_2^{NS}$, and $g_1^{NS}$. Under this assumption fermion number conservation has to be restored for $\mathbf{\Gamma}_{x\to 0}^{-}(N, a_s)$. We approach this problem in two ways numerically. In a first set of calculations we subtract a corresponding term $\propto \delta(1-x)$ from the splitting functions $P^-$, eq. (2), in each order in $a_s$ (the numerical results are labelled by 'A' later). In $N$–space this prescription leads to

$$\Gamma^{-}(N, a_s) \to \Gamma^{-}(N, a_s) - \Gamma^{-}(1, a_s)\,. \qquad (24)$$

Another possibility (denoted by 'B' in the following) is the restoration of fermion number conservation by subleading $1/N$ terms modifying $\Gamma^-$ according to

$$\Gamma^{-}(N, a_s) \to \Gamma^{-}(N, a_s) \cdot (1 - N)\,. \qquad (25)$$

The difference of the results obtained by these two procedures gives an indication on the degree of dominance of the leading terms included in the present resummation vs. uncalculated subleading contributions. Our two prescriptions for restoring fermion number conservation are analogous to the procedure in the second reference in [2] with respect to energy–momentum conservation in the unpolarized singlet case.

Before we come to the quantities studied numerically, we have to specify the input parton densities for $F^{\pm}(N, a_0)$ in (21). We choose $Q_0^2 = 4 \text{ GeV}^2$ and $\Lambda_{\overline{MS}}(N_f = 4) = 230 \text{ MeV}$. In the present study we use the same input densities and value of $\Lambda_{QCD}$ for the NLO and the resummed calculations. Specifically, in the unpolarized case we take the non–singlet combinations from the MRS(A) global fit [13]. For later use we note that $xu_v(x, Q_0^2)$ behaves $\sim x^{0.54}$ at small $x$. In the polarized case, we employ $\Delta u_v$ and $\Delta d_v$ at $Q^2 = 10 \text{ GeV}^2$ from ref. [14][5]. These densities were obtained from the DFLM [15] unpolarized valence input distributions by multiplication with $x$–dependent factors, yielding $x \Delta u_v \sim x^{0.83}$ at small $x$. We have evolved these distribution back in NLO from $Q^2 = 10 \text{ GeV}^2$ to our input scale $Q_0^2 = 4 \text{ GeV}^2$.

We are now ready to present the resummation effects on the most important non–singlet combinations. In the unpolarized case, we consider the evolution of the '–'-combination

$$\frac{1}{2} \left[ x F_3^{\nu N}(x, Q_0^2) + x F_3^{\bar{\nu} N}(x, Q_0^2) \right] = c_{F_3}^{-}(x, Q_0^2) \otimes [xu_v + xd_v](x, Q_0^2) \tag{26}$$

for an isoscalar target $N$, and of the '+'-quantity

$$F_2^{ep}(x, Q_0^2) - F_2^{en}(x, Q_0^2) = c_{F_2}^{+}(x, Q_0^2) \otimes \frac{1}{3} \left[ xu_v - xd_v + 2(x\bar{d} - x\bar{u}) \right](x, Q_0^2) \ . \tag{27}$$

In the polarized case we investigate

$$g_1^{ep}(x, Q_0^2) - g_1^{en}(x, Q_0^2) = c_{g_1}^{-}(x, Q_0^2) \otimes \frac{1}{6} \left( \Delta u_v - \Delta d_v \right)(x, Q_0^2) \ . \tag{28}$$

All these quantities are related to sum rules of phenomenological interest. $xF_3$ and $g_1$ are involved in the Gross–Llewellyn-Smith and Bjorken sum rules, respectively, which are used for the determination of $\alpha_s$. The $F_2$ difference leads to the Gottfried sum rule, which provides information on the isospin asymmetry of the light quark sea in the proton. In all these cases, experimental data have to be extrapolated towards small $x$, and hence the small-$x$ $Q^2$–evolution is of interest in principle. A '+'-combination in the polarized case is the interference structure function $g_3^{\gamma Z}$, but this quantity is hardly measureable and the effects of the resummation are even smaller than in $g_1^{NS}$. We would like to note that the effects due to resummation in the '–'-combinations do not contribute to the first moment due to fermion number conservation.

In figure 1, the NLO results for $xF_3$ and the resummed corrections $\Delta x F_3^{A,B} \equiv x F_3^{A,B} - x F_3^{NLO}$ are displayed. The resummation effects turn out to be about 1% or smaller over the whole $x$ and $Q^2$ range considered. The inclusion of subleading terms to restore the fermion number conservation (curves 'B') reduces the effect by a factor of three or more at small $x$, showing that even at $x$ as small as $x = 10^{-5}$ the resummed terms do not dominate. With respect to the size of the resummation effect, the situation is the same for $F_2^{ep} - F_2^{en}$ as shown in figure 2. For this non–singlet combination there is no constraint due to fermion number conservation.

The evolution of the polarized non–singlet structure function $g_1^{ep} - g_1^{en}$ is depicted in figure 3. Here the effect of the resummed Kernel (13) is bigger and reaches about 15% at $x = 10^{-5}$ for the $Q^2$–values considered in the figure. This enhancement is due to the flatter small-$x$ behaviour of the input densities ($\Delta u_v \sim x^{-0.17}$ vs. $u_v \sim x^{-0.46}$)[6]. Note, however, that in the $g_1^p$–range accessible for polarized electron *and* proton scattering at HERA [16] the effect is again about 1% or less. The measurement of the difference $g_1^p - g_1^n$ will have even larger statistical errors than those of $g_1^p$ alone, so that the effect due to resummation will not be resolvable in practice. With respect to the dominance of the resummed terms, the situation is the same as in the unpolarized case discussed for $xF_3$.

---

[5]We use $x_0 = 0.75$ in eq. (12) of ref. [14].

## 5 Conclusions

We have investigated the effect of the resummation of terms of order $\alpha_s^{l+1} \ln^{2l} x$ given in ref. [7] on the small-$x$ behaviour of non–singlet structure functions for deep inelastic (polarized) lepton scattering both off unpolarized and polarized targets. The comparison with terms obtained in the same order by complete NLO calculations shows the equivalence of both approaches in this limit up to order $\alpha_s^2$. Since the coefficient functions up to two–loop order for the non–singlet combinations considered contain only terms less singular in $\ln x$, the contributions to $\propto a_s^3 \ln^4 x$ in the three–loop splitting functions $P^\pm(x, a_s)$ can be predicted on the basis of ref. [7].

The numerical analysis shows that the all–order resummation of the terms $\mathcal{O}(\alpha_s^{l+1} \ln^{2l} x)$ leads only to corrections on the level of 1% for non–singlet structure functions accessible experimentally at present or in the foreseeable future. Moreover, for the '−' non–singlet combinations fermion number conservation has to be obeyed. This can lead to a further reduction of the effect by a factor of about three ore more in the small-$x$ range. The large sensitivity of the results on the prescription to implement the fermion number conservation constraint indicates the importance of so far uncalculated subleading terms $\mathcal{O}(\alpha_s^{l+1} \ln^{2l-1} x)$ down to the lowest values of $x$ considered here, $x = 10^{-5}$. If it would be possible to measure the combination $g_1^p - g_1^n$ at high precision down to $x$–values of this order, an enhancement relative to the NLO result ranging up to about 5 − 15 % were obtained, where again the spread in the correction very roughly accounts for yet unknown subleading terms.

**Acknowledgement** We would like to thank R. Kirschner, L. Lipatov, W. van Neerven, and W. Vogelsang for useful discussions. This work was supported in part by the German Federal Ministry for Research and Technology under contract No. 05 6MU93P.

---

[6]More recent parametrizations of polarized parton distributions show a similar or steeper small-$x$ behaviour, see e.g. refs. [17], leading to similar or smaller resummation effects.

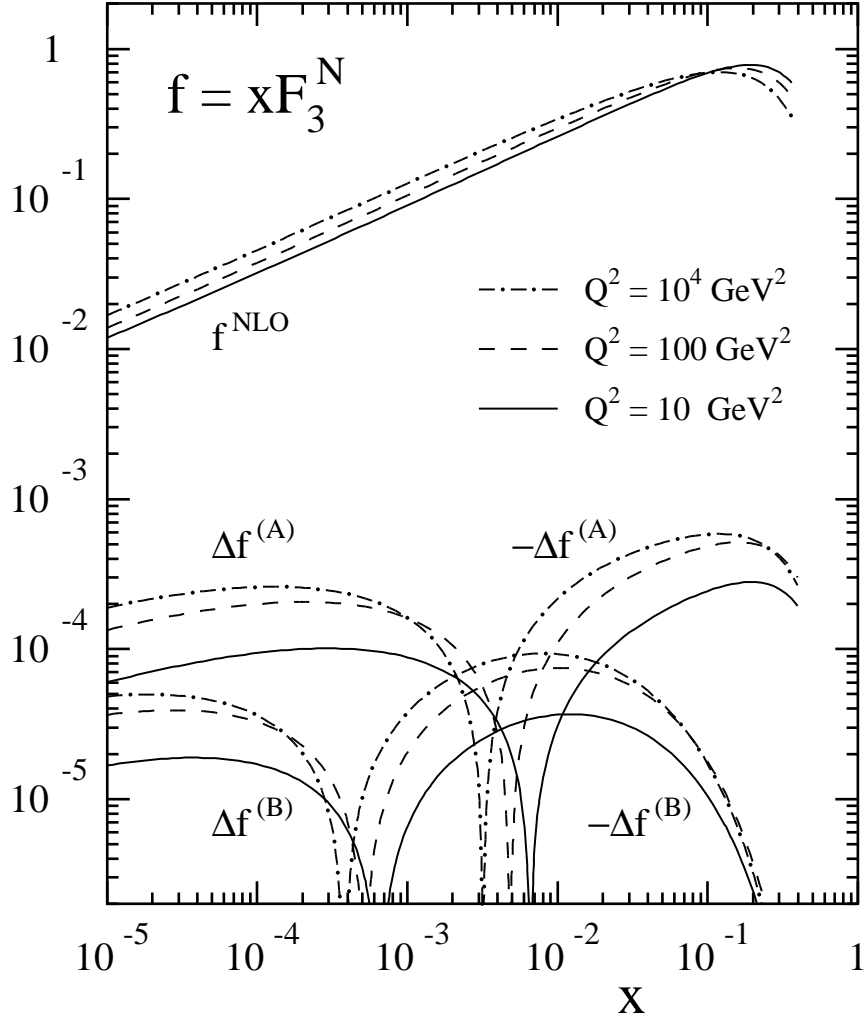

**Figure 1:** The small-$x$ $Q^2$–evolution of the non–singlet structure function $xF_3^N \equiv \frac{1}{2}(xF_3^{\nu N} + xF_3^{\bar{\nu} N})$ for an isoscalar target $N$ in NLO and the corrections to these results due to the resummed kernels derived from ref. [7]. 'A' and 'B' denote the two prescriptions for implementing the fermion number conservation discussed in the text.

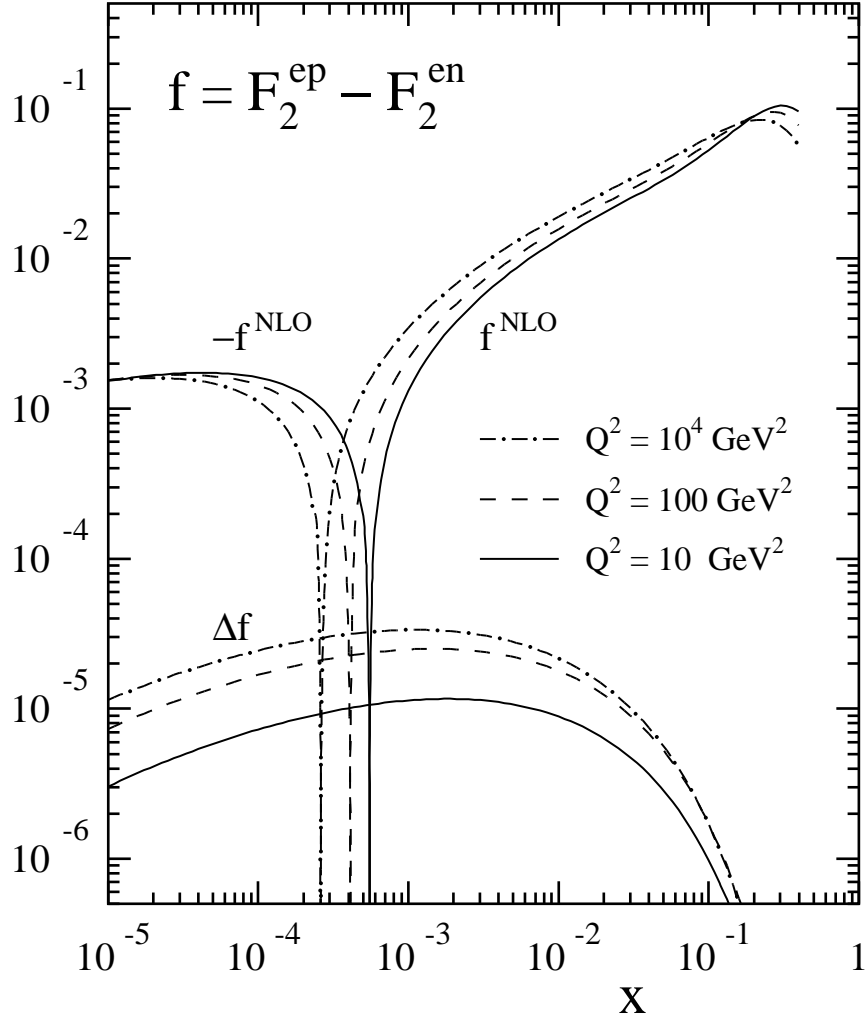

Figure 2: The same as in Fig. 1, but for the combination $F_2^{ep} - F_2^{en}$. For this '+'-combination, there is no constraint on the kernels from fermion number conservation.

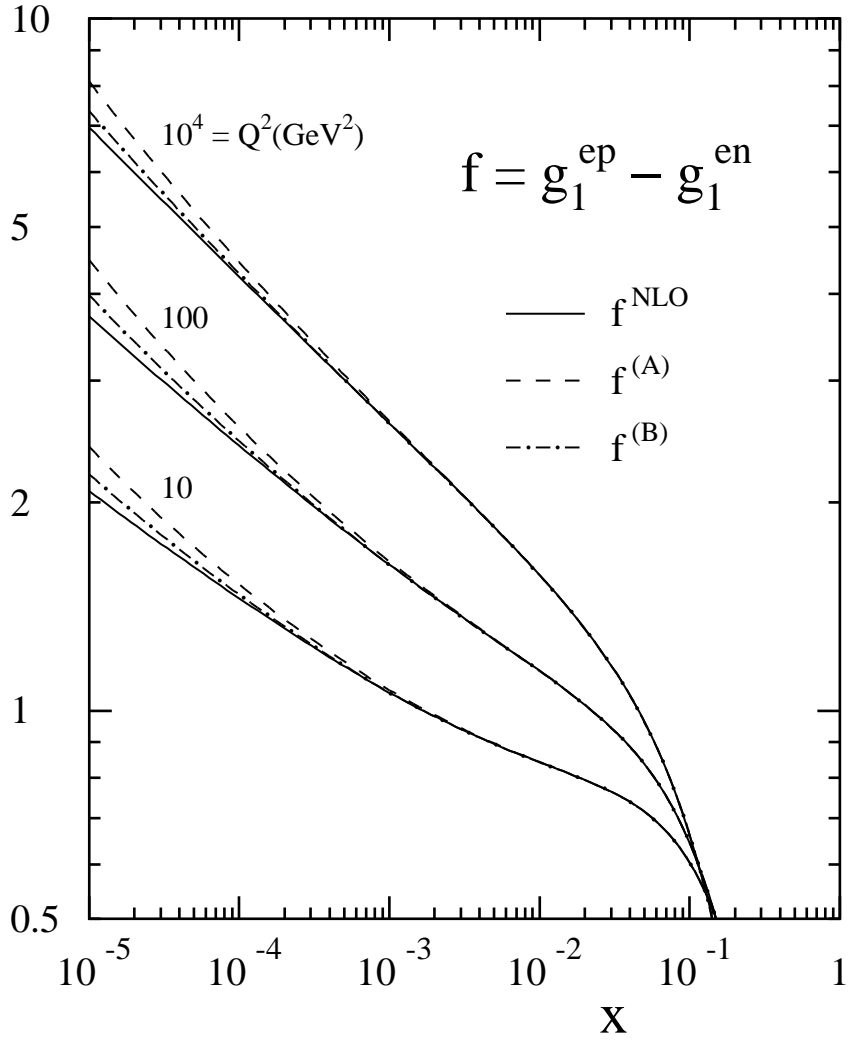

Figure 3: The small-$x$ $Q^2$–evolution of the non–singlet polarized structure–function difference $g_1^{ep} - g_1^{en}$ in NLO and with the resummed kernels taken into account. Again 'A' and 'B' denote the two prescriptions for implementing the fermion number conservation discussed in the text.